\documentclass[preprint,12pt,number,5p,twocolumn]{elsarticle}
\usepackage[utf8]{inputenc}
\usepackage{float}
\usepackage{graphics}

\usepackage[footnotesize]{caption2} 

\usepackage{threeparttable} 
\usepackage{multirow}
\usepackage{ulem}
\usepackage{flushend} 
\usepackage{verbatim}
\usepackage{graphicx}
\usepackage{color}
\usepackage{amsmath,amssymb}
\usepackage{amsmath}
\usepackage{hyperref}

\begin{document}
\title{Charge radii of $^{11-16}$C, $^{13-17}$N and $^{15-18}$O 
determined from their charge-changing cross-sections 
and the mirror-difference charge radii}

\author[b,a,c]{J.W. Zhao}
\author[a]{B.-H. Sun\corref{cor}}
\ead{bhsun@buaa.edu.cn}
\author[a,d]{I. Tanihata}
\author[a]{J.Y. Xu}
\author[b,kyz]{K.Y. Zhang}
\author[ap]{A. Prochazka}
\author[a]{L.H. Zhu}
\author[a]{S. Terashima}
\author[b]{J. Meng\corref{cor}}
\ead{mengj@pku.edu.cn}
\author[a]{L.C. He}
\author[a]{C.Y. Liu}
\author[a]{G.S. Li}
\author[els]{C.G. Lu}
\author[a]{W.J. Lin}
\author[wpl]{W.P. Lin}
\author[els,zl]{Z. Liu}
\author[wpl]{P.P Ren}
\author[els]{Z.Y. Sun}
\author[a]{F. Wang}
\author[a]{J. Wang}
\author[a]{M. Wang}
\author[els]{S.T. Wang}
\author[a]{X.L. Wei}
\author[els]{X.D. Xu}
\author[a]{J.C. Zhang}
\author[a]{M.X Zhang}
\author[els]{X.H. Zhang}

\cortext[cor]{Corresponding authors} 
\address[b]{State Key Laboratory of Nuclear Physics and Technology, School of Physics, Peking University, Beijing, 100871, China}
\address[a]{School of Physics, Beihang University, Beijing, 100191, China}
\address[c]{GSI Helmholtzcentre for Heavy Ion Research GmbH, Darmstadt, 64291, Germany}
\address[d]{RCNP, Osaka University, Ibaraki Osaka 567-0047, Japan}

\address[kyz]{Institute of Nuclear Physics and Chemistry, China Academy of Engineering Physics, Mianyang, 621900, China}
\address[ap]{Medaustron, 2700 Wiener Neustadt, Austria}
\address[els]{Institute of Modern Physics, Chinese Academy of Sciences, Lanzhou, 730000, China}
\address[wpl]{Key Laboratory of Radiation Physics and Technology of the Ministry of Education, Institute of Nuclear Science and Technology, Sichuan University, Chengdu, 610064, China}
\address[zl]{School of Nuclear Science and Technology, University of Chinese Academy of Sciences, Beijing, 100049, China}
\date{Sep. 2024}

\begin{abstract}
	Charge-changing cross-sections (CCCSs) of $^{11-16}$C, $^{13-17}$N and $^{15-18}$O on a carbon target have been determined at energies around 300 MeV/nucleon. 
 A nucleon separation energy-dependent correction factor has been introduced to the Glauber model calculation for extracting the nuclear charge radii from the experimental CCCSs. The charge radii of $^{11}$C, $^{13,16}$N and $^{15}$O thus were determined for the first time. With the new radii, we studied the experimental mirror-difference charge radii ($\Delta R_{\text {ch}}^{\text {mirror}}$) of $^{11}$B-$^{11}$C, $^{13}$C-$^{13}$N, $^{15}$N-$^{15}$O, $^{17}$N-$^{17}$Ne pairs for the first time. \textcolor{black}{We find that the $\Delta R_{\text {ch}}^{\text {mirror}}$ values of $^{13}$C-$^{13}$N and $^{15}$N-$^{15}$O pairs follow well the empirical relation to the isospin asymmetry predicted by the $ab$ $initio$ calculations, while $\Delta R_{\text {ch}}^{\text {mirror}}$ of $^{11}$B-$^{11}$C and $^{17}$N-$^{17}$Ne pairs deviate from such relation by more than two standard deviations.}
\\ 

\end{abstract}

\begin{frontmatter}
\end{frontmatter}

\section{Introduction}
\label{intro}

The nuclear charge distribution is one of the fundamental quantities for understanding nuclear structure \textcolor{black}{and essential to study the neutron skin thickness of heavy nuclei together with the neutron density distribution, which} provides insights into the equation of state for the nucleonic matter\textcolor{black}{~\cite{Roca11}}. Mirror nuclei comprise the same total number of protons and neutrons, but the number of protons in one equals the number of neutrons in the other. Properties of mirror nuclei provide access to the charge symmetry of the nuclear force and the nuclear isospin symmetry. For instance, precise charge radii data of mirror nuclei are extremely valuable for studying isospin effects in nuclei and model developments~\cite{Rein22}. The mirror-difference charge radii, $\Delta R_{\text {ch}}^{\text{mirror}}(^{A}Y/X)= R_{\text{ch}}(^{A}_{N}Y_{Z})-R_{\text {ch}}(^{A}_{Z}X_{N})$ \textcolor{black}{with $N$ ($Z$) being the neutron (proton) number of the proton-rich mirror partner}, 
for the mirror nuclei with mass number $6 \leq A \leq 56$ are predicted to have a linear relation to the isospin asymmetry ($I \equiv (N-Z)/A$) using the \textit{ab initio} coupled cluster (CC) and auxiliary field diffusion Monte Carlo (AFDMC) methods~\cite{Mirror23}. 
As an analog of the neutron skin, $\Delta R_{\text {ch}}^{\text{mirror}}$ has also been suggested to constrain the slope of the symmetry energy at saturation density~\cite{Wang13}. Test of this hypothesis has been made with different theories and sparse data~\cite{Wang13,Brown17,Yang18,Brow20,Rein22,Xu22,Huang23,Bano23,Gaut24}. However, the proper treatment of theoretical parameters (e.g., pairing correlations~\cite{Huang23,Ma24}, deformation~\cite{Ding23} and shell effects~\cite{Mohammed23}, etc.) still needs to be verified with experimental data of both bound and weakly bound nuclei over a broad range of isospin asymmetry.

When extending from stable to short-lived exotic nuclei, the electron scattering method has been precluded by the difficulty in preparing the target material, but the recent internal target-forming technique opens the possibility for electron scattering off short-lived unstable nuclei in the near future~\cite{Maza08,Suda17,Tsukada17}.
The isotope shift method has been booming in recent years~\cite{Block20}. It relies on theoretical calculations to disentangle the nuclear size-related field shift from the mass shift and can hardly access nuclei in the region containing carbon and oxygen isotopes~\cite{Yang23}. Both methods use electromagnetic probes. From the hadronic probe side, the nuclear matter distribution has been extensively studied by measuring the interaction cross sections. Analogously, the charge-changing cross section (CCCS), describing the probability of an incident nucleus losing its proton(s) by interacting with the target, is considered as an alternative method to investigate the proton distributions in nuclei~\cite{Yama11}. Such a method can in principle be applied to any isotope in the nuclear chart.

Investigations have been done to correlate CCCSs of nuclei to their root-mean-square ($rms$) point-proton distribution radii ($R_ \text p$) within the framework of the Glauber model~\cite{Meng02, AB04}. Only the collision of projectile protons with the target nucleus is considered in such models, while the projectile neutrons are treated as spectators in the charge-changing reaction. Recently, progress has been made in further understanding the charge-changing reaction mechanism. It is pointed out that the projectile neutrons also contribute to the CCCS. Such contribution has been taken into account by considering the charged-particle evaporation process (CPEP) induced by the projectile neutron removal~\cite{Tanaka21,zhao23}. The CPEP contribution to the CCCS exhibits an isospin dependence and closely correlates with the neutron-to-proton separation energy~\cite{zhao23}. This also justifies the newly developed empirical method of deducing $R_ \text p$ from the CCCSs~\cite{zhang23}, where the authors found the ratios of experimental data to the Glauber model calculations follow excellent linearity with the nucleon separation energies of the nuclide of interest. This then allows them to develop a novel approach to extract $R_ \text p$ from CCCSs  \textcolor{black}{measured at around 900 MeV/nucleon} on both hydrogen and carbon targets. \textcolor{black}{However, a remaining question is whether such an approach is fragile with the reaction energy. One of the primary goals of this work is to answer this question.}

\begin{table}[h!]
\centering
\caption{$R_\text {ch}$ and $R_\text p$ of $^{11-16}$C, $^{13-17}$N and $^{15-18}$O extracted from their experimental CCCSs, which include the ones measured in the present work \textcolor{black}{({\textit {i.e.}}, CCCSs of $^{12-16}$C, $^{14-17}$N and $^{16}$O)}, published in our previous paper~\cite{zhao23} and Ref.~\cite{Yama11}. 
}
\label{tab1}
\begin{threeparttable}
\begin{tabular}{p{0.6cm} p{1.5cm} p{1.6cm} p{1.6cm} p{1.6cm}}
\hline
  & Energy (MeV/u) & CCCS (mb) &$R_ \text p$ (fm)& $R_ \text {ch}$ (fm)\\
\hline
 $^{11}$C & 319 (38) & 716 (20) \tnote{1} & 2.18 (12) & 2.32 (11)   \\
 $^{12}$C & 228 (9) & 723 (23) & 2.41 (24) & 2.54 (22)   \\
          & 294 (5)  & 731 (52) \tnote{1} & 2.40 (26) & 2.53 (25)   \\
 $^{13}$C & 231 (8) & 720 (25) & 2.28 (21) & 2.41 (20)   \\
          & 322 (30) & 729 (22) \tnote{1} & 2.30 (13) & 2.42 (13)   \\
 $^{14}$C & 234 (7)  & 707 (13) & 2.33 (21) & 2.45 (20)  \\
          & 339 (12) & 732 (22) \tnote{1} & 2.38 (13) & 2.50 (13)  \\
          & 287 & 731 (5) \tnote{2} & 2.36 (10) & 2.48 (9)   \\
 $^{15}$C & 236 (7) & 749 (19) & 2.44 (21) & 2.56 (20)   \\
          & 327 (16) & 758 (56) \tnote{1} & 2.44 (28) & 2.56 (27)  \\
          & 285 & 743 (6) \tnote{2} & 2.36 (9)  & 2.47 (8)   \\
 $^{16}$C & 237 (6) & 738 (17) & 2.46 (22) & 2.57 (21)   \\
          & 284 & 726 (6) \tnote{2} & 2.33 (10) & 2.44 (10)   \\
\hline
 $^{13}$N & 310 (29) & 752 (35) \tnote{1} & 2.23 (17) & 2.37 (16)  \\
 $^{14}$N & 223 (7)  & 843 (32) & 2.47 (19) & 2.59 (18)  \\
          & 289 (4)  & 878 (77) \tnote{1} & 2.62 (31) & 2.74 (30)  \\
 $^{15}$N & 226 (6) & 808 (15) & 2.46 (17) & 2.58 (16)   \\
          & 315 (21) & 815 (11) \tnote{1} & 2.47 (9)  & 2.59 (9)  \\
 $^{16}$N & 227 (3) & 860 (21) & 2.56 (16) & 2.67 (15)   \\
          & 322 (18) & 813 (9)  \tnote{1} & 2.38 (7)  & 2.50 (7)  \\
 $^{17}$N & 236 (3) & 809 (17) & 2.46 (17) & 2.57 (16)   \\
          & 328 (13) & 790 (11) \tnote{1} & 2.35 (9)  & 2.47 (9)  \\
\hline
 $^{15}$O & 301 (24) & 880 (18) \tnote{1} & 2.57 (10) & 2.69 (9)  \\
 $^{16}$O & 219 (5) & 862 (17) & 2.61 (20) & 2.73 (19)   \\
          & 288 & 852 (17)\tnote{2} & 2.54 (11) & 2.66 (10)   \\
 $^{17}$O & 308 (12) & 866 (11) \tnote{1} & 2.51 (8)  & 2.62 (8)  \\
          & 294 & 896 (9) \tnote{2} & 2.61 (8)  & 2.72 (7)   \\
 $^{18}$O & 368 (2)  & 887 (39) \tnote{1} & 2.66 (17) & 2.76 (17)  \\
          & 299 & 891 (10)\tnote{2} & 2.67 (9)  & 2.78 (9)   \\
 \hline
\end{tabular}
\begin{tablenotes}
\item[1] Data \textcolor{black}{measured with the same setup, already published in Ref.~\cite{zhao23}.} 
\item[2] Data from Ref.~\cite{Yama11}.
\end{tablenotes}
\end{threeparttable}
\end{table}

\textcolor{black}{The present paper reports CCCSs of $^{12-16}$C, $^{14-17}$N and $^{16}$O on carbon measured at energies around 230 MeV/nucleon. The experiment was performed at the External Target Facility (ETF), located alongside the third focus (F3) of the second Radioactive Ion Beam Line (RIBLL2) at HIRFL-CSR~\cite{Xia02}. Secondary beams were produced by the fragmentation of $^{18}$O projectiles at 280 MeV/nucleon interacting with a beryllium target. Fragments were separated in ﬂight by the ﬁrst half (F0-F2) of RIBLL2 and then delivered to ETF. Nuclei of interest were identiﬁed event-by-event by using the magnetic rigidity ($B\rho$), time-of-ﬂight (TOF), and energy loss ($\Delta E$) information. A 2.77 $\text g/ \text {cm}^2$ thick carbon plate was used as the reaction target at ETF. The CCCSs were measured with the transmission method, where we counted the numbers of the incident and reaction particles without losing protons ({\textit {i.e.}}, $Z$-unchanged particles). To account for reactions in materials other than the target, such as detectors, we measured without the target using the same setup. The CCCS is then calculated using the equation $\sigma_\text{cc}^\text{exp} = -(1/t) \text {ln}(\gamma/\gamma_0)$, where $\gamma$ and $\gamma_0$ are the ratios of $Z$-unchanged particles for the cases with and without the carbon reaction target, respectively. $t$ is the number of target nuclei per unit area. More experimental and data analysis details can be found in previous papers~\cite{zhao23,Wang23,Sun18}.} 

\textcolor{black}{
Combining these new data with the CCCSs at around 300 MeV/nucleon measured with the same setup~\cite{zhao23}, we have identified an empirical correlation between the ratios of experimental cross sections to the theoretical ones and nucleon separation energies. This allowed us to determine the charge radii of $^{11}$C, $^{13,16}$N, $^{15}$O for the first time and investigate the mirror-difference charge radii of $^{11}\text{B}-^{11}$C, $^{13}\text C-^{13}$N, $^{15}\text N-^{15}$O, $^{17}\text N - ^{17}$Ne and $^{18} \text O-^{18}$Ne pairs.} 

\section{Results and discussions}
\label{resdis}
CCCSs of $^{11-16}$C, $^{13-17}$N and $^{15-18}$O measured in this work and published in Ref.~\cite{zhao23} are summarized in Table~\ref{tab1}. CCCSs of $^{14-16}$C and $^{16-18}$O measured at a similar energy region~\cite{Yama11} are listed as well for comparison. Our results are in good agreement with the ones from the previous experiment~\cite{Yama11}.

\begin{figure}[tbh]
	\centering
  \includegraphics[width=0.48\textwidth]{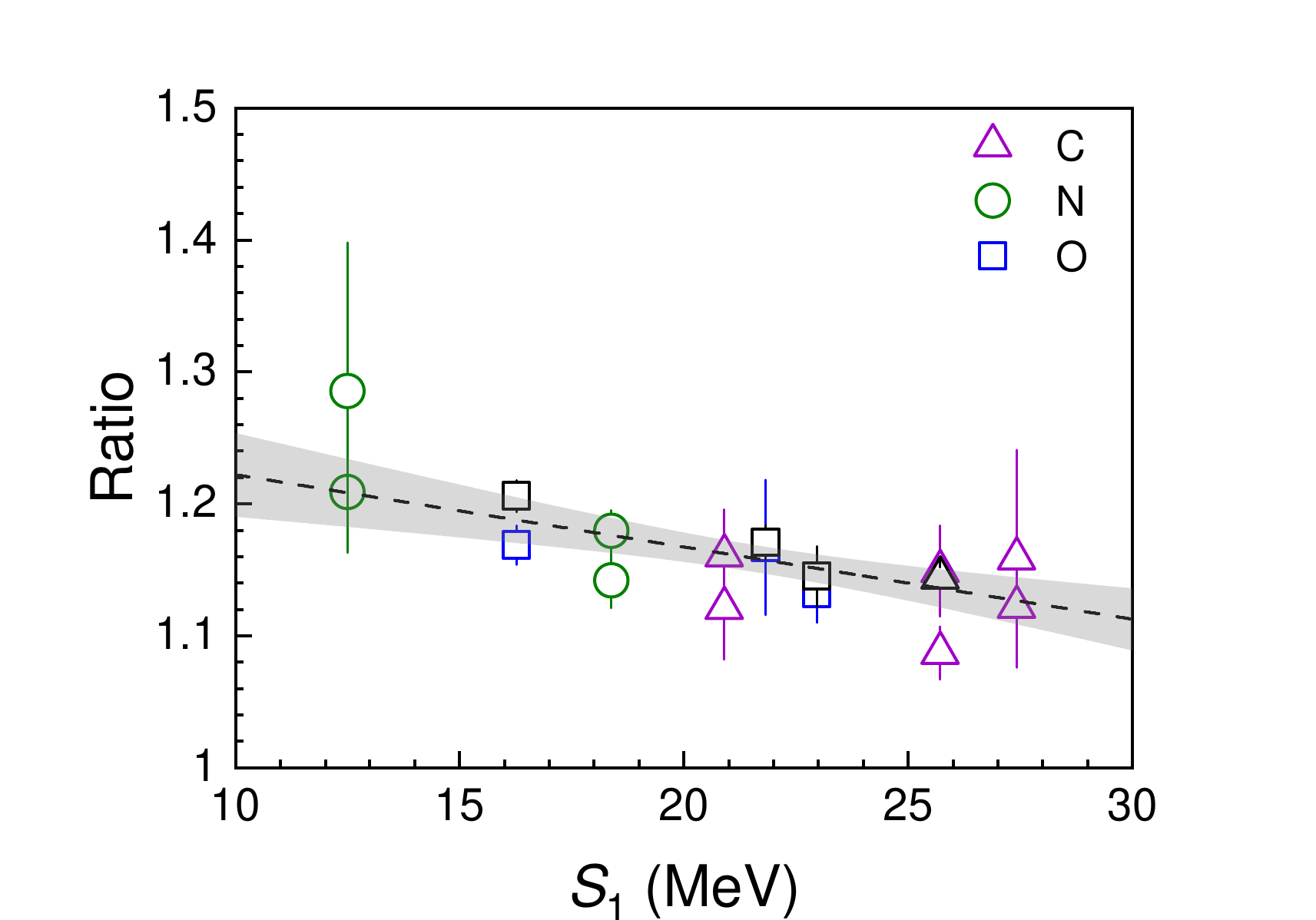}
	\caption{Ratios of experimental charge-changing cross sections to the theoretical ones, as a function of the separation energy (\textit{i.e.}, 
 $S_1 (^{A}_{Z}X_{N}) \equiv S_ \text n (^{A}_{Z}X_{N})+ S_\text p (^{A-1}_{Z}Y_{N-1}$) for $^{12-14}$C, $^{14-15}$N and $^{16-18}$O. Data are from the present work (colored markers) and Ref. ~\cite{Yama11} (black markers). The dashed line with a slope of -0.005 and an intercept of 1.28, represents the best linear fit to the data. The gray band shows the 95\% confidence level.}
	\label{f_ratio}
\end{figure}

Following the empirical method developed in Ref.~\cite{zhang23}, we have examined the ratios of experimental CCCSs to the theoretical ones as a function of the separation energy, \textit{i.e.}, $S_1 (^{A}_{Z}X_{N}) \equiv S_ \text n (^{A}_{Z}X_{N})+ S_\text p (^{A-1}_{Z}Y_{N-1}$), 
for $^{12-14}$C, $^{14-15}$N and $^{16-18}$O. Here, 
$S_ \text n (^{A}_{Z}X_{N})$ is the one-neutron separation energy of the nucleus $^{A}_{Z}X_{N}$ 
and $S_\text p (^{A-1}_{Z}Y_{N-1})$
is the one-proton separation energy of the one-neutron removed pre-fragment $^{A-1}_{Z}Y_{N-1}$. 
Ratios of the experimental CCCSs ($\sigma_\text{cc}^\text{exp}$) to the zero-range optical-limit approximation Glauber model calculations (ZRGM) ($\sigma_\text{cc}^\text{ZRGM}$),  can be well described by 
$\sigma_\text{cc}^\text{exp}/\sigma_\text{cc}^\text{ZRGM} = 
- 0.005(1) \times S_1 + 1.28(3)$ as shown in Fig.~\ref{f_ratio}. 
The slope and intercept parameters are almost identical to those for the 900 MeV/nucleon data~\cite{zhang23}. \textcolor{black}{This can be attributed to the fact that the incident energy can be well-considered in the ZRGM. Such consistent results} indicate that the correlation between $\sigma_\text{cc}^\text{exp}/\sigma_\text{cc}^\text{ZRGM}$ and $S_1$ is robust for the $p$-shell nuclei. 
\textcolor{black}{The incident energy independent feature of this correlation strengthens its application in determining the charge radius from the charge-changing reaction studies.} \textcolor{black}{As pointed out in Refs.~\cite{Tanaka21,zhao23}, the discrepancy of the Glauber model prediction from the experimental CCCS can be resolved by considering the CPEP induced by the projectile neutron removal. The projectile neutrons are removed in the first step by interacting with the target nucleons in such considerations. If the pre-fragment is highly excited, it can decay by emitting charged particles in the second step. Such CPEP will contribute to the experimental CCCS. In the first step, single neutron removal is dominant, and the probability closely relates to the one neutron separation energy (\textit{i.e.}, $S_ \text n (^{A}_{Z}X_{N})$). In the second step, single proton emission is often dominant, and the probability closely relates to the one proton separation energy of the pre-fragment (\textit{i.e.}, $S_\text p (^{A-1}_{Z}Y_{N-1})$). If both separation energies are high,  the CPEP will be suppressed. Thus, the $\sigma_\text{cc}^\text{exp}/\sigma_\text{cc}^\text{ZRGM}$ ratio dependence on the $S_1 (^{A}_{Z}X_{N})$ can be qualitatively understood~\cite{zhang23}.}

Introducing the above correction factor
to the ZRGM allows us to determine $R_\text{p}$ of nuclei with unknown radii from the experimental CCCSs. 
The point-proton density distribution of $p$-shell nuclei is assumed in the harmonic oscillator type. Table~\ref{tab1} summarizes the determined $R_ \text p$ of $^{11-16}$C, $^{13-17}$N and $^{15-18}$O. 
The error bars of $R_ \text p$ include both the systematic uncertainty from the scaling method and the statistical uncertainty. 
In the cases where CCCS statistical errors are more than 2\%, they account for more than 60\% of the uncertainties of the final $R_ \text p$. Only in the cases where CCCS statistical errors are smaller than 1\%, the scaling method starts to dominate the uncertainties of $R_ \text p$.

\begin{figure}[h!]
	\centering
	  \includegraphics[width=0.46\textwidth]{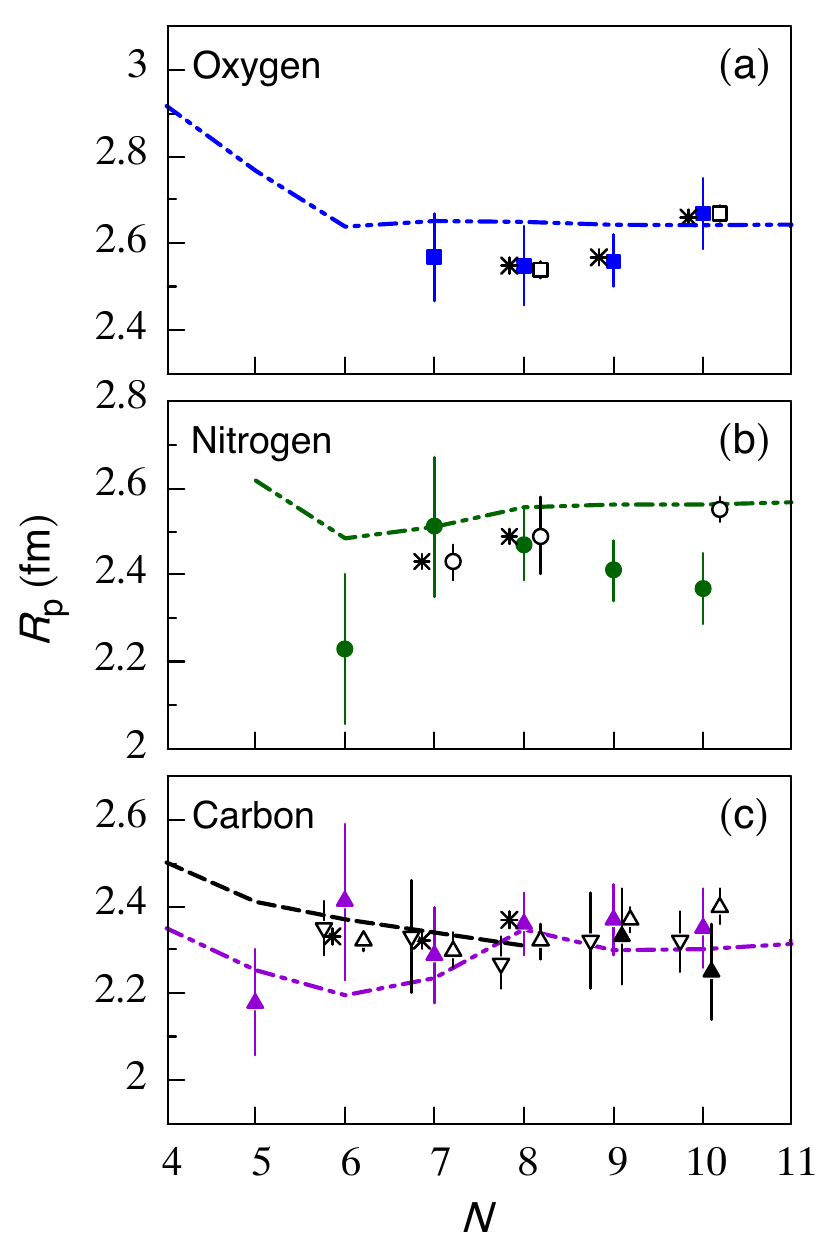}
	\caption{$R_ \text p$ determined from the present work (colored solid markers), Ref.~\cite{Yama11} (black solid markers), Refs.~\cite{Ritu16, Tran16, GSI19N, GSI22O} (open markers) in comparison with the ones (shown by $\ast$) from the electron scattering method. The data points are partially shifted in the x-axis to avoid overlapping. The dash-double-dotted lines represent predictions from the DRHBc theory~\cite{Zhou10,DRHBc,Zhang20,Zhang22,Guo24}. The dashed line shows predictions from the $ab$ $initio$ no-core shell-model theory for $^{10-14}$C~\cite{chou23}.}
	\label{fig_r}
\end{figure}

In Fig.~\ref{fig_r}, we plot $R_ \text p$ of $^{11-16}$C, $^{13-17}$N and $^{15-18}$O determined in this work, the ones from the electron scattering method~\cite{Angeli13} and previous CCCS measurements whenever available\textcolor{black}{~\cite{Yama11,Ritu16, Tran16, GSI19N, GSI22O}}. The predictions from deformed relativistic Hartree-Bogoliubov theory in continuum (DRHBc) \cite{Zhou10,DRHBc,Zhang20,Zhang22,Guo24} and 
from the \textit{ab initio} no-core shell-model~\cite{chou23} are shown for comparisons. 
\textcolor{black}{Within the present uncertainties, the new data of $^{15}$O indicates a smooth $R_ \text p$ transition over the doubly magic $^{16}$O.} While the smaller $R_ \text p$ of $^{13}$N than that of $^{14}$N may indicate an appearance of a local minimum at $N$ = 6 in the nitrogen isotopic chain. Both cases agree with the DRHBc predictions. 
The $ab$ $initio$ no-core shell-model theory and the DRHBc predict a decreasing trend of the $R_ \text p$ from $^{11}$C to $^{12}$C, but the $R_ \text p$ of $^{11}$C hints differently. \textcolor{black}{It is worth noting that $^{11}$C shows the smallest root-mean-square matter radius among that of $^{9-20}$C~\cite{Ozawa01}.}

As listed in Tabel~\ref{tab1}, we computed the charge radii ($R_ \text {ch}$) from the $R_ \text p$ using the formula,  $R_{\text {ch}}^{2}=R_{\text p}^{2}+\textlangle r_{\text p}^{2}\textrangle+(N/Z) \textlangle r_{\text n}^{2}\textrangle+\textlangle r_{\text{DF}}^{2}\textrangle$. Here, the $\textlangle r_{\text p}^{2}\textrangle = 
\textcolor{black}{(0.8409(4))^{2}} \text { fm}^{2} $ is the charge radius squared of the proton
~\cite{PDG24}, the $\textlangle r_{\text n}^{2} \textrangle = 
\textcolor{black}{-0.1155(17)} \text { fm}^{2}$ is the mean-square charge radius of the neutron
~\cite{PDG24}, and the Darwin-Folday term $\textlangle r_{\text{DF}}^{2}\textrangle = 0.033 \text { fm}^{2}$~\cite{Friar97}. 

\begin{figure}[tbh]
	\centering
 	\includegraphics[width=0.48\textwidth]{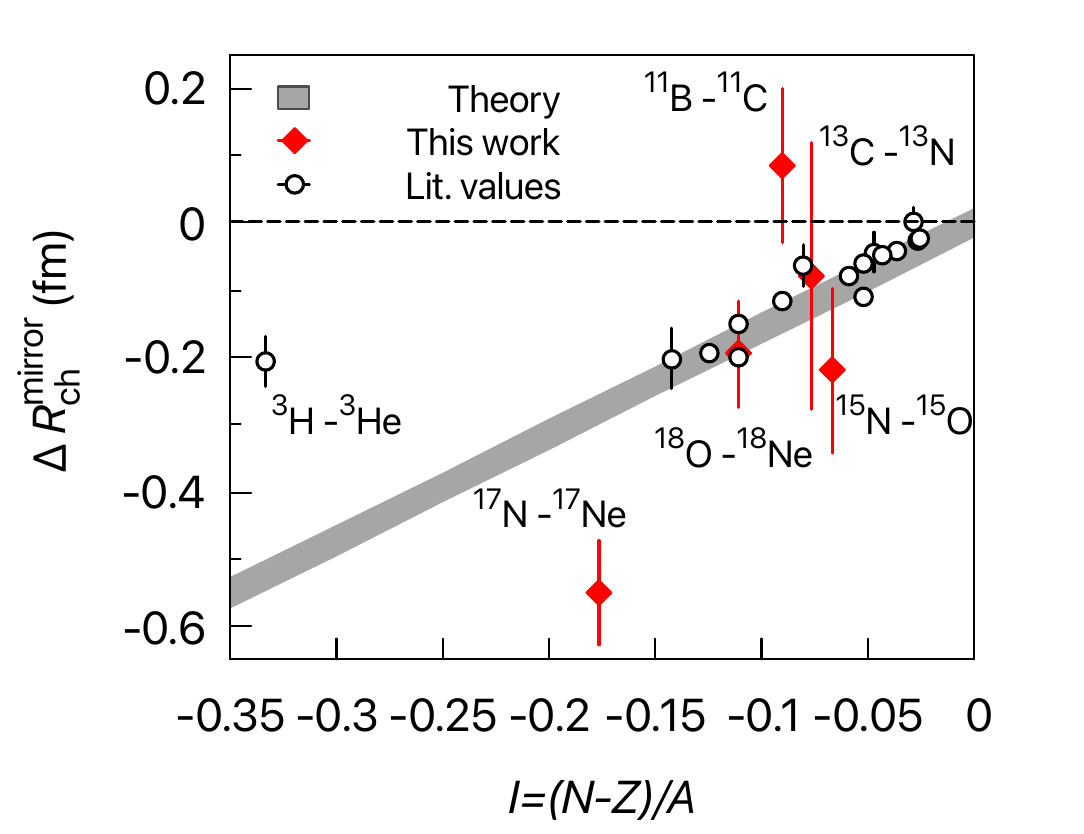}
	\caption{Mirror-difference charge radii plotted against the isospin asymmetry. Experimental results from this work (filled diamonds) are compared with the ones from literature (open circles) and the gray band representing $\Delta R_{\text {ch}}^{\text {mirror}}=1.574\cdot I \pm 0.021$.}
	\label{fig_mirror}
\end{figure}

Together with the known charge radii of $^{11}$B, $^{17,18}$Ne from Ref.~\cite{Angeli13}, we deduce the $\Delta R_{\text {ch}}^{\text {mirror}}$ of $^{11}\text{B}-^{11}$C, $^{13}\text C-^{13}$N, $^{15}\text N-^{15}$O, $^{17}\text N - ^{17}$Ne and $^{18} \text O-^{18}$Ne pairs. \textcolor{black}{$\Delta R_{\text {ch}}^{\text {mirror}}$ is obtained by subtracting $R_{\text {ch}}$ of the proton-rich mirror partner from the neutron-rich one, \textit{e.g.}, $\Delta R_{\text {ch}}^{\text {mirror}} (^{17}\text N - ^{17}$Ne) = $R_{\text {ch}} (^{17}\text {N}) - R_{\text {ch}} (^{17}\text {Ne})$}. Figure~\ref{fig_mirror} plots these results from this work and $\Delta R_{\text {ch}}^{\text {mirror}}$ of other 18 mirror nuclei pairs as a function of the isospin asymmetry, $I$.
Charge radii of $^{3}$H, $^{3}$He, $^{7}$Li, $^{7}$Be, $^{18}$O, $^{19}$F, $^{18,19,21,22}$Ne, $^{21,23}$Na, $^{34,36}$S, $^{35,37}$Cl, $^{32,34,35,37,38}$Ar, $^{39}$K, $^{39}$Ca, $^{54}$Fe from Ref.~\cite{Angeli13} and charge radii of $^{22,23}$Mg from Ref.~\cite{Yordanov12}, $^{32}$Si from Ref.~\cite{KK24}, $^{37}$K from Ref.~\cite{K21}, $^{36,37,38}$Ca from Ref.~\cite{Miller19}, $^{54}$Ni from Ref.~\cite{Pine21} are used for calculating the relevant $\Delta R_{\text {ch}}^{\text {mirror}}$.
\textcolor{black}{$\Delta R_{\text {ch}}^{\text {mirror}}$($^{18}$O$- ^{18}$Ne) determined in this work is consistent with the value deduced from the precisely known $R_{\text {ch}}$~\cite{Angeli13}. As shown in Fig.~\ref{fig_mirror}, $R_{\text {ch}}$ of the proton-rich mirror partner is always larger than that of the neutron-rich one, except for $^{11}\text B - ^{11}$C. In general, $\Delta R_{\text {ch}}^{\text {mirror}}$ exhibits a linear correlation to the isospin asymmetry $I$ as the one (\textit {i.e.}, $R_{\text {ch}}^{\text {mirror}}=1.574\cdot I \pm 0.021$) identified in the $ab$ $initio$ calculation~\cite{Mirror23}, where the experimental charge radii of individual mirror nuclei are reproduced within a few percent. 
\textcolor{black}{A deviation of $\Delta R_{\text {ch}}^{\text {mirror}}$ from such correlation implies an anomaly in the nuclear structure contributing to the isospin symmetry breaking within the mirror pairs.} 
For instance, the deviation of $\Delta R_{\text {ch}}^{\text {mirror}} (^{17}\text N - ^{17}$Ne) is probably due to the proton halo character of $^{17}$Ne~\cite{17Ne08,17Ne}. The more extreme case of $\Delta R_{\text {ch}}^{\text {mirror}} (^{3}\text H - ^{3}$He) is not surprising since the structure effects (\textit {e.g.}, pairing correlation, short-range-correlation~\cite{Li22}) are more pronounced for such light nuclei with large isospin asymmetry. $\Delta R_{\text {ch}}^{\text {mirror}} (^{11}\text B - ^{11}$C) slightly deviates from that linear correlation. However, a recent laser spectroscopic study on $^{10,11}\text B$ ~\cite{11B19} has pointed out that the nuclear charge radius of $^{11}\text B$ from Ref.~\cite{Angeli13} suffers from large uncertainties.}
As shown in Fig.~\ref{fig_mirror}, we have extended the experimental $\Delta R_{\text {ch}}^{\text {mirror}}$ down to the isospin asymmetry of -0.176. Our experimental data, including both the bound ({\textit {e.g.}}, $^{15}$N) and weakly bound proton-rich mirror partners, provide ideal test cases for the treatment of shell effects and pairing correlation in the nuclear theories~\cite{Rein22}.     

\section{Summary}
\label{sum}
In the present study, we reported the charge radii of $^{11-16}$C, $^{13-17}$N, and $^{15-18}$O \textcolor{black}{determined from the relevant CCCSs on a carbon target at around 300 MeV/nucleon. For the first time, we have identified a robust correlation of the experimental cross section relative to the models for $p$-shell nuclei at this energy range.}  
The correction factor is almost identical to those found at 900 MeV/nucleon data \textcolor{black}{indicating that the reaction energy dependence has been eliminated by taking the ratio of the experimental CCCS to the Glauber model prediction. This method opens a new vista of determining nuclear charge radii from the charge-changing reaction studies.}
With the $R_ \text p$ of $^{11}$C, $^{13,16}$N, and $^{15}$O measured for the first time, we show the $R_ \text p$ of oxygen isotopes show a smooth transition over the magic neutron number $N$ = 8 and suggest a local minimum at $N$ = 6 in the nitrogen isotopes. The smaller proton radius of $^{11}$C than that of $^{12}$C tends to differ from the predictions of the \textit{ab initio} no-core shell model and the DRHBc theories. 

These new data allow us to deduce the experimental mirror-difference charge radii $\Delta R_{\text {ch}}^{\text {mirror}}$ of $^{11}$B-$^{11}$C, $^{13}$C-$^{13}$N, $^{15}$N-$^{15}$O, $^{17}$N-$^{17}$Ne pairs for the first time. \textcolor{black}{$\Delta R_{\text {ch}}^{\text {mirror}}$ of $^{13}$C-$^{13}$N and $^{15}$N-$^{15}$O pairs are consistent with the \textit{ab initio} theoretical predictions. The deviation of $\Delta R_{\text {ch}}^{\text {mirror}}$ of $^{17}$N-$^{17}$Ne coincides with the proton halo structure of $^{17}$Ne. The slight deviation of $\Delta R_{\text {ch}}^{\text {mirror}}$ of $^{11}$B-$^{11}$C implies the uncertainty in the present knowledge on the charge radius of $^{11}$B.} Although present data do not have high precision, an improved statistic in the CCCS experiment will help to quantify such linearity better. \textcolor{black}{Experimental $\Delta R_{\text {ch}}^{\text {mirror}}$ will also be valuable in revealing distinct nuclear ($e.g.$, structure) effects contributing to the isospin symmetry breaking and testing the treatment of shell effects and pairing correlation in the nuclear theories.}

\section*{Acknowledgments}
We thank the HIRFL-CSR accelerator team for their efforts to provide a stable beam condition during the experiment. This work was supported partially by the National Natural Science Foundation of China (Grant Nos. 12325506, 11961141004, 11922501, and 12135004), the ``111 Center" (Grant No. B20065), the Strategic Priority Research Program of Chinese Academy of Sciences (Grant No. XDB34010000), and the National Key R\&D Program of China (Grant No. 2016YFA0400504, 2018YFA0404402). J. Z. acknowledges the support by the Office of China Postdoctoral Council (OCPC) under the International Postdoctoral Exchange Fellowship
Program (No. 20191034).

\biboptions{sort&compress}


\begin{thebibliography}{00}

\bibitem{Roca11} X. Roca-Maza, M. Centelles, X. Viñas, and M. Warda, \href{https://doi.org/10.1103/PhysRevLett.106.252501}{Phys. Rev. Lett. \textbf{106}, 252501 (2011).}

\bibitem{Rein22} P. Reinhard and W. Nazarewicz, \href{https://doi.org/10.1103/PhysRevC.105.L021301}{Phys. Rev. C \textbf{105}, L021301 (2022).}

\bibitem{Mirror23} S. J. Novario, D. Lonardoni , S. Gandolfi and G. Hagen, \href{https://doi.org/10.1103/PhysRevLett.130.032501}{Phys. Rev. Lett. \textbf{130}, 032501 (2023).}

\bibitem{Wang13} N. Wang and T. Li, \href{http://dx.doi.org/10.1103/PhysRevC.88.011301}{Phys. Rev. C \textbf{88}, 011301 (2013).}

\bibitem{Brown17} B.A. Brown, \href{https://doi.org/10.1103/PhysRevLett.119.122502}{Phys. Rev. Lett. \textbf{119}, 122502 (2017).}

\bibitem{Yang18} J. Yang and J. Piekarewicz, \href{https://doi.org/10.1103/PhysRevC.97.014314}{Phys. Rev. C \textbf{97}, 014314 (2018).}

\bibitem{Brow20} B.A. Brown, K. Minamisono, J. Piekarewicz, et al., \href{https://doi.org/10.1103/PhysRevResearch.2.022035}{Phys. Rev. Research \textbf{2}, 022035 (2020).}

\bibitem{Xu22} J. Xu, Z. Li, B.-H. Sun, et al., \href{https://doi.org/10.1016/j.physletb.2022.137333}{Phys. Lett. B \textbf{833}, (2022) 137333.}

\bibitem{Huang23} Y.N. Huang, Z.Z. Li, and Y.F. Niu, \href{https://doi.org/10.1103/PhysRevC.107.034319}{Phys. Rev. C \textbf{107}, 034319 (2023).}

\bibitem{Bano23} P. Bano, S.P. Pattnaik, M. Centelles, et al., \href{https://doi.org/10.1103/PhysRevC.108.015802}{Phys. Rev. C \textbf{108}, 015802 (2023).}

\bibitem{Gaut24} S. Gautam, A. Venneti, S. Banik and B.K. Agrawal, \href{https://doi.org/10.1016/j.nuclphysa.2024.122832}{Nucl. Phys. A \textbf{1043}, 122832 (2024).}

\bibitem{Ma24} Xiao-Rong Ma, Shuai Sun, Rong An, Li-Gang Cao, \href{ https://iopscience.iop.org/article/10.1088/1674-1137/ad47a8}{Chinese Phys. C \textbf{48}, 084104 (2024).}

\bibitem{Ding23} Meng-Qi Ding, Ping Su, De-Qing Fang, Si-Min Wang, \href{ https://iopscience.iop.org/article/10.1088/1674-1137/ace680}{Chinese Phys. C \textbf{47}, 094101 (2023).}

\bibitem{Mohammed23} Ruqaya A. Mohammed and Wasan Z. Majeed, \href{ https://doi.org/10.1134/S106377882302014X}{Physics of Atomic Nuclei, \textbf{86}, 77 (2023).}

\bibitem{Maza08}
X. Roca-Maza, M. Centelles, F. Salvat, and X. Viñas, \href{https://doi.org/10.1103/PhysRevC.78.044332}{Phys. Rev. C \textbf{78}, 044332 (2008).}

\bibitem{Suda17}
T. Suda and H. Simon, \href{https://doi.org/10.1016/j.ppnp.2017.04.002}{Prog. Part. Nucl. Phys. \textbf{96}, 1 (2017).}

\bibitem{Tsukada17}
K. Tsukada, A. Enokizono, T. Ohnishi et al., \href{https://doi.org/10.1103/PhysRevLett.118.262501}{Phys. Rev. Lett. \textbf{118}, 262501 (2017).}

\bibitem{Block20}
M. Block, M. Laatiaoui and S. Raeder, \href{https://doi.org/10.1016/j.ppnp.2020.103834}{Prog. Part. Nucl. Phys. \textbf{116}, 103834 (2020).}

\bibitem{Yang23}
X.F. Yang, S.J. Wang S.G. Wilkins and R.F. Garcia Ruiz, \href{https://doi.org/10.1016/j.ppnp.2022.104005}{Prog. Part. Nucl. Phys. \textbf{129}, 104005 (2023).}

\bibitem{Yama11} T. Yamaguchi, I. Hachiuma, A. Kitagawa et al., \href{https://doi.org/10.1103/PhysRevLett.107.032502}{Phys. Rev. Lett. \textbf{107}, 032502 (2011).}

\bibitem{Meng02} J. Meng, S.G. Zhou, I. Tanihata, \href{https://doi.org/10.1016/S0370-2693(02)01574-5}{Phys. Lett. B \textbf{532}, 209 (2002).}

\bibitem{AB04} A. Bhagwa, Y.K. Gambhir, \href{https://doi.org/10.1103/PhysRevC.69.014315}{Phys. Rev. C \textbf{69}, 014315 (2004).}

\bibitem{Tanaka21} M. Tanaka, M. Takechi, A. Homma, A. Prochazka,
M. Fukuda, D. Nishimura, et al., \href{https://doi.org/10.1103/PhysRevC.106.014617}{Phys. Rev. C \textbf{106}, 014617 (2022).}

\bibitem{zhao23} J.W. Zhao, B.-H. Sun, I. Tanihata, et al., \href{https://doi.org/10.1016/j.physletb.2023.138269}{Phys. Lett. B \textbf{847}, (2023) 138269.}

\bibitem{zhang23} J.-C. Zhang, B.-H. Sun, I. Tanihata, et al., \href{https://doi.org/10.1016/j.scib.2024.03.051}{Sci. Bull. \textbf{69}, (2024) 1647.}

\bibitem{Xia02} J.W. Xia et al., \href{https://doi.org/10.1016/S0168-9002(02)00475-8}{Nuclear Instruments and Methods in Physics Research A 488 (2002) 11.}

\bibitem{Wang23} C.-J. Wang, G. Guo, H.J. Ong, Y.-N. Song, B.-H. Sun, I. Tanihata, et al., \href{ https://doi.org/10.1088/1674-1137/acd366}{Chinese Phys. C \textbf{47}, 084001 (2023).}

\bibitem{Sun18} B.-H. Sun, J.-W. Zhao, X.-H. Zhang, L.-N. Sheng, Z.-Y. Sun, I. Tanihata,
\textit{et al.}, \href{ https://doi.org/10.1016/j.scib.2017.12.005} {Science Bulletin \textbf{63}, 78  (2018).}

\bibitem{Angeli13} I. Angeli and K.P. Marinova, \href{https://doi.org/10.1016/j.adt.2011.12.006}{Atomic Data and Nuclear Data Tables \textbf{99}, 69 (2013).} 

\bibitem{GSI19N} S. Bagchi, R. Kanungo, W. Horiuchi et al., \href{https://doi.org/10.1016/j.physletb.2019.01.024}{Phys. Lett. B \textbf{790}, 251 (2018).}

\bibitem{chou23} P. Choudhary, P.C. Srivastava, M. Gennari, and P. Navr\'atil, \href{https://doi.org/10.1103/PhysRevC.107.014309}{Phys. Rev. C \textbf{107}, 014309 (2023).}

\bibitem{Zhou10} S.G. Zhou, J. Meng , P. Ring, and E.G Zhao, \href{https://doi.org/10.1103/PhysRevC.82.011301}{Phys. Rev. C \textbf{82}, 011301(R) (2010).}

\bibitem{DRHBc} L.L Li, J. Meng, P. Ring, E.-G. Zhao, S.-G. Zhou, \href{https://doi.org/10.1103/PhysRevC.85.024312}{Phys. Rev. C \textbf{85}, 024312 (2012).}

\bibitem{Zhang20} K. Zhang, M.K. Cheoun, Y.B. Choi, P.S. Chong,
J. Dong, L. Geng, et al., \href{https://doi.org/10.1103/PhysRevC.102.024314}{Phys. Rev. C \textbf{102}, 024314 (2020).} 

\bibitem{Zhang22} K. Zhang, M.K. Cheoun, Y.B. Choi, P.S. Chong,
J. Dong, X. Du, et al., \href{https://doi.org/10.1016/j.adt.2022.101488}{Atomic Data and Nuclear Data Tables \textbf{144}, 101488 (2022).} 

\bibitem{Guo24} Peng Guo, Xiaojie Cao, Kangmin Chen, Zhihui Chen, Myung-Ki Cheoun, Yong-Beom Choi, et al., \href{https://doi.org/10.1016/j.adt.2024.101661}{Atomic Data and Nuclear Data Tables \textbf{158}, 101661 (2024).} 

\bibitem{Ritu16} R. Kanungo, W. Horiuchi, G. Hagen  et al., \href{https://doi.org/10.1103/PhysRevLett.117.102501}{Phys. Rev. Lett. \textbf{117},102501 (2016).}

\bibitem{Tran16} D.T. Tran, H.J. Ong, T.T. Nguyen et al., \href{https://doi.org/10.1103/PhysRevC.94.064604}{Phys. Rev. C \textbf{94}, 064604 (2016).}

\bibitem{GSI22O} S. Kaur,  R. Kanungo, W. Horiuchi et al., \href{https://doi.org/10.1103/PhysRevLett.129.142502}{Phys. Rev. Lett. \textbf{129}, 142502 (2022).}


\bibitem{Ozawa01} A. Ozawa, O. Bochkarev, L. Chulkov, et al., \href{https://doi.org/10.1016/S0375-9474(01)00563-2}{Nucl. Phys. A \textbf{691}, 599 (2001).}


\bibitem{PDG24} Particle data group Collaboration, \href{https://pdg.lbl.gov/index-2024.html}{Particle data group, as of August 4th (2024).}



\bibitem{Friar97} J.L. Friar, J. Martorell, and D.W.L. Sprung, \href{https://doi.org/10.1103/PhysRevA.56.4579}{ Phys. Rev. A \textbf{56}, 4579 (1997)}

\bibitem{Yordanov12} D.T. Yordanov, M.L. Bissell, K. Blaum, et al.,\href{https://doi.org/10.1103/PhysRevLett.108.042504}{Phys. Rev. Lett. \textbf{108}, 042504 (2012).}

\bibitem{KK24} Kristian König, Julian C. Berengut, Anastasia Borschevsky, et al.,\href{https://doi.org/10.1103/PhysRevLett.132.162502}{Phys. Rev. Lett. \textbf{132}, 162502 (2024).}

\bibitem{K21} Á. Koszorús, X.F. Yang, W.G. Jiang, et al., \href{https://doi.org/10.1038/s41567-020-01136-5}{Nat. Phys. \textbf{17}, 439–443 (2021).} 

\bibitem{Miller19} A.J. Miller, K. Minamisono, A. Klose, et al., \href{https://doi.org/10.1038/s41567-019-0416-9}{Nat. Phys. \textbf{15}, 432–436 (2019).} 

\bibitem{Pine21} S.V. Pineda, K. König, D.M. Rossi, et al., \href{https://doi.org/10.1103/PhysRevLett.127.182503}{Phys. Rev. Lett. \textbf{127}, 182503 (2021).}

\bibitem{17Ne08} W. Geithner, T. Neff, G. Audi, et al.,\href{https://doi.org/10.1103/PhysRevLett.101.252502}{Phys. Rev. Lett. \textbf{101}, 252502 (2008).}

\bibitem{17Ne} C. Lehr, F. Wamers, F. Aksouh et al., \href{https://doi.org/10.1016/j.physletb.2022.136957}{Phys. Lett. B \textbf{827}, 136957 (2022).}

\bibitem{Li22} S. Li, R. Cruz-Torres, N. Santiesteban, et al., \href{https://doi.org/10.1038/s41586-022-05007-2}{Nat. Phys. \textbf{609}, 41–45 (2022).} 

\bibitem{11B19} Bernhard Maaß, Thomas Hüther, Kristian König, et al., \href{https://doi.org/10.1103/PhysRevLett.122.182501}{Phys. Rev. Lett. \textbf{122}, 182501 (2019).}



\end{thebibliography}
\end{document}